\documentclass[
aps,pre,reprint, amsmath, amssymb,superscriptaddress
]{revtex4-2}
\usepackage{graphicx} 
\usepackage{braket}
\usepackage[colorlinks,linkcolor=red,urlcolor=blue,citecolor=blue]{hyperref}
\usepackage{xcolor}
\usepackage[bb=boondox]{mathalfa}

\newcommand{\Tr}[1]{\mathrm{Tr}\left[#1\right]}

\newcommand{\dt}[1]{\frac{\mathrm{d}#1}{\mathrm{d}t}}

\newcommand{\w}[2]{w_{#1}^{#2}}

\begin{document}
\title{Information geometry approach to quantum stochastic thermodynamics}
\author{Laetitia P. Bettmann}
\email{bettmanl@tcd.ie}
\affiliation{School of Physics, Trinity College Dublin, College Green, Dublin 2, D02K8N4, Ireland}
\author{John Goold}
\email{gooldj@tcd.ie}
\affiliation{School of Physics, Trinity College Dublin, College Green, Dublin 2, D02K8N4, Ireland}
\affiliation{Trinity Quantum Alliance, Unit 16, Trinity Technology and Enterprise Centre, Pearse Street, Dublin 2, D02YN67, Ireland}
\begin{abstract}
Recent advancements have revealed new links between information geometry and classical stochastic thermodynamics, particularly through the Fisher information (FI) with respect to time. Recognizing the non-uniqueness of the quantum Fisher metric in Hilbert space, we exploit the fact that any quantum Fisher information (QFI) can be decomposed into a metric-independent incoherent part and a metric-dependent coherent contribution. We demonstrate that the incoherent component of any QFI can be directly linked to entropic acceleration, and for GKSL dynamics with local detailed balance, to the rate of change of generalized thermodynamic forces and entropic flow, paralleling the classical results.
Furthermore, we tighten a classical uncertainty relation between the geometric uncertainty of a path in state space and the {time-averaged} rate of information change and demonstrate that it also holds for quantum systems. We generalise a classical geometric bound on the entropy rate for far-from-equilibrium processes by incorporating a non-negative quantum contribution that arises from the geometric action due to coherent dynamics. 
Finally, we apply an information-geometric analysis to the recently proposed quantum-thermodynamic Mpemba effect, demonstrating this framework's ability to capture thermodynamic phenomena.
\end{abstract}
\date{\today}
\maketitle
\section{Introduction}
Stochastic thermodynamics is vital for understanding the dynamic behavior of systems that deviate from thermal equilibrium during physical processes~\cite{seifert_stochastic_2012, sekimoto_stochastic_nodate}.  Key achievements in the field include the development of fluctuation theorems, such as the Jarzynski equality~\cite{jarzynski_nonequilibrium_1997} and the Crooks fluctuation theorem~\cite{crooks_nonequilibrium_1998}. Both provide exact relations between equilibrium and non-equilibrium quantities and impose constraints on entropy production in non-equilibrium processes. Another breakthrough is the formulation of thermodynamic uncertainty relations~\cite{seifert_stochastic_2012, pietzonka_universal_2018, horowitz_thermodynamic_2020}, which establish trade-offs between dissipation and precision of currents in systems operating far from equilibrium. In parallel, there has been a concerted effort to extend these classical results to quantum systems, where unique features like coherence and entanglement require a more nuanced treatment. This has led to quantum generalizations of fluctuation theorems and uncertainty relations, broadening the scope of stochastic thermodynamics to encompass quantum dynamics~\cite{esposito_nonequilibrium_2009, campisi_fluctuation_2011, miller_thermodynamic_2021-1}.\\
Recently, several connections between geometric properties and thermodynamics have been uncovered. Among them is a theory known as geometric thermodynamics, which was developed for the slow-driving regime in classical and, subsequently, quantum settings~\cite{crooks_measuring_2007, sivak_thermodynamic_2012, salamon_thermodynamic_1983, scandi_quantum_2020, abiuso_geometric_2020}. As a tool, is has proven particularly useful for the optimization task of identifying protocols that minimize entropy production~\cite{rolandi_finite-time_2023, rolandi_collective_2023, zulkowski_geometry_2012, rotskoff_geometric_2017, brandner_thermodynamic_2020, li_geodesic_2022, eglinton_geometric_2022, eglinton_thermodynamic_2023, frim_geometric_2022, terren_alonso_geometric_2022, mehboudi_thermodynamic_2022}.

Furthermore, the well-established connection between information and thermodynamics~\cite{parrondo_thermodynamics_2015}, central to resolving the famous Maxwell's demon paradox~\cite{maruyama_colloquium_2009}, has recently been extended using tools from information geometry. In this framework, probability distributions are treated as points within a geometric space. A key concept here is the Fisher information~\cite{Burbea1982, Rao1992, Fisher_1922}, a fundamental statistical measure that quantifies how much information a random variable provides about an unknown parameter on which the probability distribution depends~\cite{kay_fundamentals_1993}. Its interpretation as a geometric quantity is grounded in \v{C}encov's theorem, which identifies Fisher information as the unique Riemannian metric in classical probability spaces that is contractive under arbitrary stochastic maps (noisy transformations)~\cite{cencov_statistical_2000}. Through its role as a metric, it facilitates the exploration of geometric concepts like statistical distance, divergence, curvature, and geodesics~\cite{nielsen_elementary_2020}.

Geometric analysis has shown that the Fisher information with respect to the time parameter is particularly significant in thermodynamics. It measures how much the probability distribution  at a given point in time changes along a path in the probability space traced by the time-evolving state, effectively serving as an instantaneous speed. This made it possible to link the Fisher information to entropic acceleration, providing a direct connection between geometrical properties and thermodynamic quantities for classical systems~\cite{ito_stochastic_2018, nicholson_nonequilibrium_2018}. Crucially, these results hold without the typical constraints of near-equilibrium conditions or slow driving.
In quantum systems, similar geometric analysis has successfully assigned an operational meaning to the quantum Fisher information (QFI) in quantum thermodynamics~\cite{marvian_operational_2022}. Additionally, speed limits that generalize classical results~\cite{nicholson_time-information_2020} have been established~\cite{garcia-pintos_unifying_2022, pires_generalized_2016, bringewatt_arxiv_2024, sekiguchi2024speedlimits}.

Quantum systems often require a more refined approach compared to their classical counterparts due to quantum correlations that are captured by density matrices rather than classical probability distributions. This has interesting consequences in information geometry: while the metric in classical probability spaces is unique~\cite{cencov_statistical_2000}, the states of quantum systems live within a Hilbert space where a variety of metrics can be used to measure distances between states, depending on the chosen inner product. Petz’s work on monotone metrics has been significant in this regard. He investigated the conditions under which metrics on the space of density matrices remain monotone under completely positive, trace-preserving (CPTP) maps~\cite{petz_covariance_2002, petz_introduction_2011, petz_riemannian_1996, petz_monotone_1996}. Petz identified a family of such monotone metrics, all of which are contractive under CPTP maps, ensuring that the statistical distance between quantum states does not increase through physical operations, consistent with the principle that information  cannot be increased merely by processing data~\cite{nielsen_quantum_2010}. {Notable examples include the symmetric logarithmic derivative (SLD) metric (smallest), the Kubo-Mori-Bogoliubov (KMB) metric, the harmonic mean (HM) metric (largest), and the Wigner-Yanase (WY) metric. For a comprehensive review on this topic, we refer readers to~\cite{scandi_quantum_2024}.}

The aim of our work is to extend classical results connecting information geometry and stochastic thermodynamics to quantum systems. 
To this end, we first show that the uncertainty relation connecting the geometric uncertainty in the path in state space and the {time-averaged} rate of information change, previously established for classical systems~\cite{nicholson_nonequilibrium_2018}, carries over seamlessly to the quantum domain. {In fact, we find that the bound can be tightend both in classical and quantum settings}. Provided the dynamics is arbitrary differentiable and trace-preserving, any QFI can be split into a metric-independent incoherent and a metric-dependent coherent contribution~\cite{amari_methods_2007, bengtsson_geometry_2006}.
Building on this result, we then show that any QFI can be uniquely linked to the entropic acceleration through their shared incoherent component. This allows us to demonstrate that when the quantum system's dynamics are governed by a GKSL master equation~\cite{lindblad_generators_1976, gorini_completely_1976}, the incoherent part of the QFI can be expressed in terms of both the rate of change in generalized thermodynamic forces, driving probability currents between instantaneous eigenstates of the system's density matrix, as well as the entropic flux exchanged with an environment, mirroring the established relationship in classical systems~\cite{ito_stochastic_2018}. Further, we generalise a classical geometrical bound on the rate of change of the von Neumann entropy to the quantum case by the addition of a non-negative term arising from the coherent contribution to the geometric action.
To further illustrate these findings, we analyze the recently proposed quantum thermodynamic Mpemba effect~\cite{moroder_thermodynamics_2024} through the lens of information geometry.

The paper is structured as follows: section~\ref{sec: FI} reviews the Fisher information with respect to time, emphasizing its geometric interpretation. Section~\ref{sec: QFI} extends this framework to quantum systems, introducing the family of quantum Fisher informations (QFIs) and their explicit forms for arbitrary differentiable trace-preserving quantum dynamics (section~\ref{sec: t_param}). In section~\ref{sec: info_geo_thermo_classical}, we examine classical results linking the Fisher information with respect to the time parameter to entropic acceleration, entropic flow, and entropy production rates. Section~\ref{sec: info_geo_thermo_quantum} presents our first main result, showing how these classical results extend to quantum systems. We then present our second key result, which builds on the first by extending a classical bound on the entropy rate { — }expressed in terms of geometric quantities describing the path traced by the system’s evolution in state space { — }by including a non-negative contribution from the coherent part of the geometric action. Finally, in section~\ref{sec: mpemba}, we illustrate the relationship between quantum thermodynamics and information geometry using the recently reported quantum-thermodynamic Mpemba effect~\cite{moroder_thermodynamics_2024}. We demonstrate that the relaxation speed-up is captured by the ``ratio of completion", a quantity grounded in the geometric statistical distance, and analyze the geometric uncertainty. Lastly, we examine a previously derived geometric bound on the time evolution of arbitrary observables~\cite{garcia-pintos_unifying_2022, bringewatt_arxiv_2024} with a focus on the Hamiltonian, establishing a connection to the dissipated heat.
\section{Fisher information with respect to time}
\label{sec: FI}
The concept of Fisher information arises in mathematical statistics \cite{Fisher_1922}.
Let a path be defined by the set of discrete probability distributions {over the set of discrete states $X$} that a system is described by {at any given time $t$} as it evolves over a time interval of length $\tau$.
Formally, we express this as $P(X) = \left\{p: X\to \mathbb{R}\hspace{0.2cm}|\hspace{0.2cm}0\leq p_x(t)\leq 1\hspace{0.2cm} \forall x\in X, \hspace{0.2cm}\sum_x p_x(t) = 1\right\}$, ensuring the non-negativity of the probabilities $p_x(t)$ as well as the normalisation of the {set of probability distributions} $P(X)$. We now assume that there is a finite number of control parameters $\theta(t) = (\theta_1(t),\dots, \theta_M(t) )$, that the path taken depends on. The statistical manifold, to which the path is confined, is $\Theta= \left\lbrace p(x|\theta(t)):\hspace{0.2cm}\theta(t)\right\rbrace$ with the coordinates set by the control parameters.
The metric tensor $m_{ij}$ that equips the manifold of probability distributions with a statistical measure of distance, $\mathrm{d}s^2 = \frac{1}{4}\sum_{ij} m_{ij}\mathrm{d}\theta_i \mathrm{d}\theta_j$, is the so-called Fisher matrix \cite{amari_methods_2007}
\begin{equation}
    m_{ij} = \sum_x p_x(\theta) \frac{ \partial \log p_x(\theta)}{\partial \theta_i} \frac{ \partial \log p_x(\theta)}{\partial \theta_j}.
\end{equation}
We may also take the another viewpoint --- we may consider time itself as a parameter. 
The Fisher information about the time parameter is then given by
\begin{equation}
    F(t) = \sum_{ij} \frac{\mathrm{d}\theta_i}{\mathrm{d}t} m_{ij}\frac{\mathrm{d}\theta_j}{\mathrm{d}t} = \sum_x p_x(t) \left[\frac{\mathrm{d}\log p_x(t)}{\mathrm{dt}}\right]^2.
\end{equation}
It becomes apparent that the Fisher information itself satisfies the requirements of a metric, since
\begin{equation}
    \mathrm{d}s^2 = \frac{1}{4} F(t) \mathrm{d}t^2.
\end{equation}
Interestingly, by \v{C}encov's theorem, it is in fact the only Riemannian metric on the set of probability distributions that is contractive under stochastic maps \cite{cencov_statistical_2000}. 
Since the Fisher information is a metric, it relates to the line element between two distributions infinitesimally displaced from one another on the manifold.
The path $\gamma$ that a probability distribution traces on the manifold has the { path} length \cite{wootters_statistical_1981}
\begin{equation}
    \mathcal{L} = \int_\gamma \mathrm{d}s = \frac{1}{2}\int_{0}^{\tau} \mathrm{d}t\sqrt{F(t)}.
\end{equation}
The above length is interpreted as the statistical distance in the space of probability distributions.
Note that the term $\mathrm{d}s/\mathrm{d}t = \frac{1}{2}\sqrt{F(t)}$ expresses the instantaneous speed along the path $\gamma$ at time $t$.
Further, it is useful to define the ratio of completion as the ratio between the statistical distance traced in probability space until time $t$, $\mathcal{L}(t)$, and the statistical distance at some chosen final time $\tau$, $\mathcal{L}(\tau)$, as 
\begin{equation}
    \mathcal{R}_\tau(t) = \frac{\mathcal{L}(t)}{\mathcal{L}(\tau)}.
\end{equation}
This quantity has recently proven useful in the information-geometric analysis of asymmetries of thermal processes in classical and open quantum systems \cite{tejero_asymmetries_2024, ibanez_heating_2024} and is central in our analysis of the thermodynamic quantum Mpemba effect~\cite{moroder_thermodynamics_2024} in Sec.~\ref{sec: mpemba}.
\section{Quantum extensions}
\label{sec: QFI}
Also in the context of quantum dynamics, the quantum Fisher information (QFI) with respect to the time parameter has attracted significant interest and has  proven useful, particularly in the development of quantum speed limits~\cite{marvian_operational_2022, girolami_how_2019, garcia-pintos_unifying_2022, yadin_quantum_2024, pires_generalized_2016, bringewatt_arxiv_2024}. In the following, we review the QFI and its geometric interpretation, in analogy to the classical case discussed in the previous section.

In quantum settings, the objects of interest are states represented by density matrices rather than probability distributions. Like in the classical case, a statistical line element with respect to time $t$ in quantum state space may be defined, however, the choice of the metric is not unique.
Rather, there exists a family of metrics, all interpretable as a different QFI, characterised by the Morozova, \v{C}encov and Petz theorem \cite{petz_introduction_2011, petz_riemannian_1996, petz_covariance_2002, toth_extremal_2013, morozova_markov_1991, scandi_quantum_2024}. 
{According to this characterization, }any metric contractive under stochastic evolution must yield a squared line element of the form (up to multiplication by a scalar constant) \cite{petz_introduction_2011} 
\begin{equation}
\label{eq: ds_q}
        \mathrm{d}s^2 = \frac{1}{4}  \sum_{x, y } \frac{|\mathrm{d}\hat{\rho}_{xy}|^2}{p_x f(p_y/p_x)},
\end{equation}
where $\hat{\rho}=\sum_x p_x \ket{x}\bra{x}$, {$\mathrm{d}\hat{\rho}_{xy} := \braket{x|\mathrm{d}\hat{\rho}|y}$}, and $\lbrace p_x\rbrace $ form a discrete probability distribution, and the function $f$ is 1) an operator monotone, so that for any positive semidefinite operators $A$ and $B$ such that $A \leq B$, then $f(A) \leq f(B)$,  2) self-inversive, so that $f(x) = x f(1/x)$,  and 3) normalised, meaning $f(1) = 1$. Assuming that the density operator $\hat{\rho}(t)$ depends analytically on $t$, $\mathrm{d}\hat{\rho}(t) = \partial_t \hat{\rho} \mathrm{d}t$,
\begin{equation}
\label{eq: ds_q}
    \mathrm{d}s^2 = \frac{1}{4}  \sum_{x, y } \frac{|\partial_t \hat{\rho}_{xy}(t)|^2}{p_x(t) f(p_y(t)/p_x(t))}\mathrm{d}t^2,
\end{equation}
{where $\partial_t \hat{\rho}_{xy} := \braket{x(t)|\partial_t \hat{\rho}_t|y(t)}$.} 
The above lets us identify the general form of any QFI about the parameter $t$ as
\begin{equation}
    F_Q(t) = \sum_{x, y } \frac{|\partial_t\hat{\rho}_{xy}(t)|^2}{p_x(t) f(p_y(t)/p_x(t))}.
\end{equation}
Prominent members of the QFI family are the symmetric logarithmic derivative (SLD) QFI ($f_{\mathrm{SLD}}(x) =\frac{x+1}{2}$), which is the smallest QFI, the Wigner-Yanase (WY) QFI ($f_{\mathrm{WY}}(x) = \frac{1}{4}(\sqrt{x}+1)^2$), and the harmonic mean (HM) QFI ($f_{\mathrm{HM}}(x) = \frac{2x}{x+1}$), which is the largest QFI. For further details, we refer the reader to~\cite{scandi_quantum_2024}. 

Either of the different statistical distances $\mathcal{L}_f$, since the respective metrics are contractive under quantum stochastic maps by construction, represents a faithful measure of distinguishability over the quantum state space. Additionally, one may ask the following optimisation question:  what is the geodesic path, that is the path with constant 
speed, that connects the initial state $\hat{\rho}(0)$ and the final state $\hat{\rho}(\tau)$, so that $\mathcal{L}^\mathrm{geo}_f\leq \mathcal{L}_f$. It is the closest analogue of a straight line on a curved manifold. While the protocol for traversing the state space along the geodesic path itself is often nontrivial to obtain (analytically), when using either the SLD~\cite{uhlmann_density_1993} or the WY~\cite{gibilisco_wigneryanase_2003} QFI metrics, we can make a statement about the length of the geodesic paths via closed form expressions, respectively,
\begin{align}
\label{eq: geodesic_length}
    \mathcal{L}^\mathrm{geo}_\mathrm{SLD}(\hat{\rho}_1, \hat{\rho}_2) &= \arccos{\sqrt{F(\hat{\rho}_1, \hat{\rho}_2)}} \\
    \mathcal{L}^\mathrm{geo}_\mathrm{WY}(\hat{\rho}_1, \hat{\rho}_2) &=  \arccos{A(\hat{\rho}_1, \hat{\rho}_2)},
\end{align}
where $F(\hat{\rho}_1, \hat{\rho}_2)= \left(\Tr{\sqrt{\sqrt{\hat{\rho}_1} \hat{\rho}_2\sqrt{\hat{\rho}_1}}}\right)^2$ is the Uhlmann fidelity and $A(\hat{\rho}_1, \hat{\rho}_2) = \Tr{\sqrt{\hat{\rho}_1}\sqrt{\hat{\rho}_2}}$ is called the quantum affinity.
\section{General time evolution}
\label{sec: t_param}
In the following, we assume arbitrary differentiable trace-preserving dynamics of a quantum system with respect to time $t$. We first show that any QFI with respect to the time parameter can be split into an incoherent part and a coherent part~\cite{amari_methods_2007, bengtsson_geometry_2006}. The former can then be identified as the classical Fisher information of the probability distribution, constructed from the instantaneous spectrum of the density matrix. The latter is a genuine quantum contribution that quantifies the state's coherence in the eigenbasis of some time-dependent Hermitian operator, that in some cases may be identified with the physical Hamiltonian generating unitary and thus coherent evolution.\\
If the dynamics is differentiable and trace-preserving, then the state at a given time $t$, $\hat{\rho}_t$, can be written as
\begin{equation}
\label{eq: general_evolution}
    \hat{\rho}_t = \hat{U}_t \hat{\chi}_t \hat{U}_t^\dagger.
\end{equation}
The spectral decomposition of the state at time $t$ and the initial state are, respectively, $\hat{\rho}_t = \sum_x p_x(t)\ket{x(t)}\bra{x(t)}$ and $\hat{\rho}_0 = \sum_x p_x(0)\ket{x(0)}\bra{x(0)}$. The unitary operator $U_t$ transforms between the respective eigenbasis elements, so that $\ket{x(t)} = \hat{U}_t \ket{x(0)}$. The matrix $\hat{\chi}_t$ is given by $\hat{\chi}_t = \sum_x p_x(t) \ket{x(0)}\bra{x(0)}$, i.e. it is the matrix that is diagonal in the original basis, but with the time evolved spectrum on the diagonal.
Eq.~\eqref{eq: general_evolution} simply expresses that the state might have a time-evolving spectrum (through incoherent processes) in a time-evolving eigenbasis (through coherent evolution). This allows us to define the Hermitian operator $\hat{H}_t = i(\mathrm{d}\hat{U}_t/\mathrm{d}t)\hat{U}_t^\dagger$, which may be viewed as an effective ``Hamiltonian", and the differential equation governing the evolution of $\hat{\rho}_t$ is~\cite{avron_adiabatic_1987, girolami_how_2019, alipour_shortcuts_2020}
\begin{equation}
\label{eq: drhodt}
    \partial_t \hat{\rho}_t = -i\left[\hat{H}_t, \hat{\rho}_t\right] + \hat{U}_t \partial_t \hat{\chi}_t \hat{U}_t^\dagger.
\end{equation}
Coming back to the QFI, we note that the factor common to every member of the QFI family is {the change of the matrix element $|\partial_t \hat{\rho}_{xy}|^2$ in the instantaneous eigenbasis of $\hat{\rho}_t$.} 
Let us plug in the form in Eq.~\eqref{eq: drhodt}:
\begin{equation}
\begin{split}
    &|\braket{x(t)|\partial_t \hat{\rho}_t|y(t)}|^2 \\&= |\braket{x(t)|-i\left[\hat{H}_t, \hat{\rho}_t\right]|y(t)}  + \braket{x(t)|\hat{U}_t \partial_t \hat{\chi}_t \hat{U}_t^\dagger|y(t)}|^2\\
     &=|-i\braket{x(t)|\left[\hat{H}_t, \hat{\rho}_t\right]|y(t)} + \partial_t(p_x(t)) \delta_{xy}|^2.
\end{split}
\end{equation}
Analysing the cross term $\braket{x(t)|\hat{H}_t \hat{\rho}_t -\hat{\rho}_t \hat{H}_t |y(t)} \delta_{xy} = (p_y(t)-p_x(t))\braket{x(t)|\hat{H}_t|y(t)} \delta_{xy}$, we note that it vanishes for any pair $x$ and $y$, and we find
\begin{equation}
\begin{split}
    &|\braket{x(t)|\partial_t \hat{\rho}_t|y(t)}|^2 \\
     &=(p_y(t)-p_x(t))^2|\braket{x(t)|\hat{H}_t|y(t)}|^2 +    (\partial_t p_x(t) )^2\delta_{xy}.
\end{split}
\end{equation}
Any QFI with respect to the parameter time $t$ can thus be split into an incoherent contribution $F^{IC}_Q(\hat{\rho}_t)$ and a coherent contribution $F^{C}_Q(\hat{\rho}_t)$, so that
\begin{align}
    F_Q(\hat{\rho}_t) =& F^{IC}_Q(\hat{\rho}_t)+F^{C}_Q(\hat{\rho}_t),\\
    F^{IC}_Q(\hat{\rho}_t) =& \sum_x p_x(t) \left(\frac{\mathrm{d}}{\mathrm{d}t}\log p_x(t)\right)^2, \\
    F^{C}_Q(\hat{\rho}_t) =& \sum_{x \neq y} \frac{|\partial_t \hat{\rho}_{xy}|^2}{p_x(t) f(p_y(t)/p_x(t))}.
\end{align}
Therefore, the QFI, similar to the classical FI, is sensitive to changes in the spectrum of the state via $F^{IC}_Q(\hat{\rho}_t)$. In addition, however, unlike in the classical case where the eigenbasis is fixed, the eigenbasis in quantum systems can undergo unitary rotations so that the density matrix does not commute with its time-derivative. This results in the coherent contribution $F^{C}_Q(\hat{\rho}_t)$. 

Interestingly, the coherent contribution of any of the QFI variants is a measure of coherence, also referred to as asymmetry, with respect to the time-dependent Hermitian operator $\hat{H}_t$ \cite{streltsov_colloquium_2017, marvian_operational_2022}.\\
Since the square infinitesimal length element is given by $\mathrm{d}s^2 = 1/4 \, F_Q(\hat{\rho}_t) \mathrm{d}t^2$, the statistical distance in state space is given by
\begin{equation}
    \mathcal{L} = \frac{1}{2} \int_{0}^{\tau}\sqrt{F_Q(\hat{\rho}_t)} \mathrm{d}t.
\end{equation}
We now define the statistical divergence
\begin{equation}
    \mathcal{J}= \frac{\tau}{4} \int_{0}^{\tau} F_Q(\hat{\rho}_t)  \mathrm{d}t.
\end{equation}
In Riemannian geometry $\mathcal{J}/2\tau$ is referred to as the action or energy of the path due to its similarity with the kinetic energy integral in classical mechanics (i.e. $\int_{0}^{\tau} E_\mathrm{kin} \mathrm{d}t$, with $E_\mathrm{kin} = \frac{1}{2} v^2$ and $v$ is a speed and the mass is set to unity). 
As a consequence of the Cauchy-Schwarz inequality, the statistical divergence bounds the squared statistical distance, so that $\mathcal{J}-\mathcal{L}^2\geq 0$. Equality of $\mathcal{J}$ and $\mathcal{L}^2$ is achieved only when the integrand remains constant along the path—i.e., when the speed, expressed in terms of the QFI, is constant, indicating that the system follows a geodesic trajectory.
Interestingly, it was shown in the context of classical stochastic thermodynamics, that in the quasi-static limit, the thermodynamic length and divergence encode the dissipation of finite time thermodynamic transformations \cite{salamon_thermodynamic_1983, Salamon1985, janyszek_riemannian_1989, nicholson_nonequilibrium_2018}.\\
\section{Information geometry for stochastic thermodynamics}
\label{sec: QME}
Before delving into the connection between information geometry and quantum stochastic thermodynamics, it is useful to revisit the recent classical results~\cite{ito_stochastic_2018, nicholson_nonequilibrium_2018}, which serve as the foundation for our quantum analysis. These results establish a fundamental link between information geometry and thermodynamics in classical stochastic processes governed by Markovian master equations~\cite{sekimoto_stochastic_2010, seifert_stochastic_2012, schnakenberg_network_1976, andrieux_fluctuation_2007}. To this end, we will briefly review the relationship between line elements and observables in stochastic thermodynamics, which enables the interpretation of information-geometric quantities within the framework of stochastic thermodynamics. We then show how this connection extends to the quantum domain.
\subsubsection{Classical systems}
\label{sec: info_geo_thermo_classical}
Following ~\cite{ito_stochastic_2018, nicholson_nonequilibrium_2018}, let us consider a discrete-state system with $N>1$ states. We denote the set of states as $X$. The system is assumed to be weakly coupled to one or more thermal reservoirs, with interactions inducing transitions between the system's states. The dynamics of the system can be described via the  dynamics of a time-dependent probability distribution $p_t = (p_1(t), \dots, p_N(t))$, so that $p_x(t)$ is the probability to find the system in the state labeled by $x\in X$ at any given time $t$. We further assume that the set of probabilities associated with given states $x\in X$ evolve according to a time-continuous Markovian master equation      
\begin{equation}
    \dot{p}_x(t) = \sum_y w^{xy}(t) p_y(t),
\end{equation}
where $w^{xy}(t)>0$ (if $x\neq y$) is the transition rate from state $y\to x$, and $w^{yy}(t) = -\sum_{x\neq y} w^{xy}(t)$, which leads to $\sum_x w^{xy}(t) = 0$, ensuring the normalisation of $p_t$.  
This allows us to rewrite the master equation in terms of the probability currents from state $y$ to $x$, $j^{xy}(t) = w^{xy}(t) p_y(t) - w^{yx}(t) p_x(t)$, as
\begin{equation}
     \dot{p}_x(t) = \sum_y j^{xy}(t).
\end{equation}
We impose the local detailed-balance condition, by which the stochastic entropy change in the environment $\phi^{xy}$ due to the transition from state $y$ to $x$ at time $t$ is given by 
\begin{equation}
    \phi^{xy}(t) = \log \frac{w^{xy}(t)}{w^{yx}(t)}.
\end{equation}
By Shannon, the information content associated with a given state $y$, also known as surprisal or self-information, is given by \cite{shannon_mathematical_1948}
\begin{equation}
    I_y (t) = -\log p_y(t).
\end{equation}
Accordingly, the local difference in self-information, or relative surprisal, for states $y$ and $x$ is
\begin{equation}
    I_{xy}(t) = -\log \frac{p_y(t)}{p_x(t)}.
\end{equation}
Given these notions, the time-derivative of the Shannon entropy $S(t) = -\sum_x p_x(t)\log p_x(t)$ may be identified as the current-weighted average, { denoted by $\langle \langle A \rangle \rangle = \frac{1}{2}\sum_{xy} j^{xy}(t) A_{xy}$ for any function of the transition $y\to x$}, of the surprisal~\cite{ito_stochastic_2018, esposito_three_2010}
\begin{equation}
\begin{split}
    \dot{S}(t) &= -\sum_x \dot{p}_x(t) \log p_x(t) \\
    & = -\frac{1}{2} \sum_{x,y} j^{xy}(t) I_{xy}(t) \\
    &= -\langle \langle I(t) \rangle \rangle.
\end{split}
\end{equation}
From a physically, rather than information theoretically motivated standpoint, we may, alternatively, split it into two contributions: the entropy production rate in the system $\sigma(t)$ and the entropy flux from the system to the environment $\Phi(t)$, $\dot{S}(t) = \sigma(t) - \Phi(t)$, with
\begin{align}
    \sigma(t) &= \left\langle\left \langle \log\frac{w^{xy}(t)p_y(t)}{w^{yx}(t)p_x(t)} \right\rangle \right\rangle,\\
    \Phi(t) &= \left\langle \left\langle \log\frac{w^{xy}(t)}{w^{yx}(t)} \right \rangle \right \rangle.
\end{align}
Further, intuition can be gained when introducing the generalised thermodynamic force associated with driving the probability current from state $y$ to $x$~\cite{gingrich_dissipation_2016}, given by
\begin{equation}
\label{eq: gen_force_classical}
    f^{xy}(t)=\log\left[\frac{w^{xy}(t)p_y(t)}{w^{yx}(t)p_x(t)}\right].
\end{equation}
In near-equilibrium thermodynamics macroscopic currents $J$ can be linearly related to thermodynamic forces $F$ via the macroscopic mobility $\mu$ ($J = \mu F$)~\cite{onsager_reciprocal_1931, onsager_reciprocal_1931-1}. A similar mathematical structure emerges in systems far from equilibrium, where the microscopic probability current between states $y$ and $x$, $j^{xy}$, is linearly related to the generalised force $f^{xy}$ instead via the dynamical state mobility $m^{xy}$~\cite{van_vu_thermodynamic_2023}, so that $j^{xy} = m^{xy} f^{xy}$. {Note, that this relation serves to implicitly define $m^{xy}$.}
This lets us identify the entropy production and flux as current averages 
 \begin{align}
    \sigma(t) &= \left\langle\left \langle f(t) \right\rangle\right\rangle,\\
    \Phi(t) &= \left\langle \left\langle \phi(t) \right \rangle \right \rangle.
\end{align}
The Fisher information, which may be interpreted as a measure of fluctuations in the surprisal rate,
\begin{equation}
\begin{split}
     F(t) &= \sum_x p_x(t) \left(\frac{\mathrm{d}\log p_x(t)}{\mathrm{d}t}\right)^2,
\end{split}
\end{equation}
may be alternatively expressed as a current weighted average of the change in the surprisal $F(t) = \langle\langle \dot{I}(t) \rangle \rangle$. It is worth noting here, that the surprisal rate $\dot{I}(t)$ is non-zero only in situations in which the system is out of (thermodynamic) equilibrium. More intuitively, the state's ability to encode the time parameter relies on the fact that the state evolves with time. \\
The second derivative of the Shannon entropy, the entropic acceleration, can then be expressed as~\cite{nicholson_nonequilibrium_2018}
\begin{equation}
    \ddot{S}(t) = -\frac{\mathrm{d}}{\mathrm{d}t}\langle \langle I(t) \rangle \rangle 
    = -\sum_x \ddot{p}_x(t) \log p_x(t) - F(t)
\end{equation}
 We now take a closer look at the term $ \mathcal{B} = -\sum_x \ddot{p}_x(t) \log p_x(t)$: to this end, we first note that the the rate of change of the Shannon entropy $\dot{S}(t) = -\langle\langle I(t)\rangle \rangle$ can be thought of as the rate of change of the average information. Therefore, the Shannon entropy's second derivative $\ddot{S}(t)= -\frac{\mathrm{d}}{\mathrm{d}t}\langle \langle I(t)\rangle \rangle $ in turn tells us how the rate of change of the average information changes in time --- it is the {\textit{rate of change of the current averaged relative surprisal}}. The Fisher information, however, is an average over the rate of change in information between each set of states $x$ and $y$, that is $F(t) = \langle \langle \dot{I}(t)\rangle \rangle$ --- it is the {\textit{current average over the rate of change of the relative surprisal}. Therefore, $\mathcal{B} = \left\langle \left \langle \frac{\mathrm{d}}{\mathrm{d}t}  I(t) \right \rangle \right \rangle - \frac{\mathrm{d}}{\mathrm{d}t} \left \langle \left \langle I(t) \right\rangle \right \rangle$ is the difference of the average local information rate and the bulk information rate. }
Using the definition in Eq.~\eqref{eq: gen_force_classical} one finds \cite{ito_stochastic_2018}
\begin{equation}
\begin{split}
    F(t) & = - \frac{1}{2} \sum_{x, y} j^{xy}(t) \frac{\mathrm{d}f^{xy}(t)}{\mathrm{d}t} + \frac{1}{2}  \sum_{x, y} j^{xy}(t)\frac{\mathrm{d}}{\mathrm{d}t}\log \frac{w^{xy}(t)}{w^{yx}(t)} \\
    &= -\left\langle\left\langle \frac{\mathrm{d}f(t)}{\mathrm{d}t} \right\rangle \right\rangle + \left\langle\left\langle \frac{\mathrm{d}\phi(t)}{\mathrm{d}t} \right\rangle\right \rangle.
\end{split}
\end{equation}
The above equation provides a stochastic thermodynamic interpretation of information geometry in classical Markovian discrete state systems with local detailed balance. {It holds generally, even far from equilibrium. However, noting that at equilibrium $f^{xy} = 0$ for any pair of states $x$ and $y$, we find that near equilibrium, the Fisher information simplifies to the rate of change of the entropic flux, as discussed in~\cite{ito_stochastic_2018}.} 

Furthermore, due to the positivity of the Fisher information,  $\left\langle\left\langle \frac{\mathrm{d}\phi(t)}{\mathrm{d}t} \right\rangle\right \rangle \geq \left\langle\left\langle \frac{\mathrm{d}f(t)}{\mathrm{d}t} \right\rangle \right\rangle $~\cite{ito_stochastic_2018}. The inequality is saturated only in a steady state where the Fisher information vanishes. This tells us that the variation of the thermodynamic force rate is {converted} to the environmental entropy change rate, and any mismatch between the two quantities, expressing a loss in the entropy change rate transfer, is a result of non-stationarity~\cite{ito_stochastic_2018}.
\subsubsection{Quantum generalisation}
\label{sec: info_geo_thermo_quantum}
In the following, we show that an analogous connection between the entropy and any quantum Fisher information can be drawn, assuming that the state of an open quantum system evolves under a GKSL master equation ($\hbar =1$),
\begin{equation}
    \dt{\hat{\rho}} = -i \left[\hat{H}_t, \hat{\rho}\right] + \sum_k \mathcal{D}\left[\hat{L}_k\right] \hat{\rho},
\end{equation}
where the dissipator is given by {$\mathcal{D}\left[\hat{L}\right]\hat{\rho} = \hat{L}\hat{\rho} \hat{L}^\dagger - \lbrace \hat{L}^\dagger \hat{L}, \hat{\rho}  \rbrace/2$}, {satisfying local detailed balance~\cite{Horowitz_2013, manzano_quantum_2018}. Under this condition, the jump operators must come in pairs $(k^\prime, k)$ that fulfill
\begin{equation}
    \hat{L}_k = e^{\phi^k/2}\hat{L}_k^\prime,
\end{equation}
where $\phi^k = -\phi^{k^\prime}$ is the entropy change in the environment upon the action of the jump operator $\hat{L}_k$.
Let us express the state $\hat{\rho}_t$ in its instantaneous eigenbasis $\hat{\rho}_t = \sum_x p_x(t)  \ket{x(t) }\bra{x(t) }$.
Crucially, we may then define transition rates $w_k^{xy}(t)  = |\braket{x(t) |\hat{L}_k|y(t) }|^2$~\cite{Funo2019, van_vu_thermodynamic_2023}.\\}
Note here, that the Hamiltonian $\hat{H}_t$ generally does not coincide with the Hermitian time-dependent operator  introduced in section~\ref{sec: t_param}, but rather can be reconstructed from the eigenstates of the time-evolved state $\hat{\rho}_t$ from which $\hat{U}_t$ is obtained.\\
 When taking the derivative of $p_x(t)  = \braket{x(t) |\hat{\rho}_t |x(t) }$, one finds a classical master equation for the instantaneous spectrum~\cite{Funo2019, van_vu_thermodynamic_2023}
\begin{equation}
    \dt{p_x(t) } = \sum_{k, y (y\neq x)} \left[\w{k}{xy}(t) p_y(t)  - \w{k}{yx}(t) p_x(t) \right].
\end{equation}
Further, let us define
\begin{align}
    j_k^{xy}(t)  & = \w{k}{xy}(t)  p_y(t)  - \w{k^\prime}{yx}(t)  p_x(t), \\
    f_k^{xy}(t)  &= \log\frac{\w{k}{xy}(t)  p_y(t) }{\w{k^\prime}{yx}(t)  p_x(t) },\\
    \phi_k^{xy}(t) &=  \log\left(\frac{w^{xy}_k(t)}{w^{yx}_{k^\prime}(t)}\right),
\end{align}
where $j_k^{xy}(t)$ is the probability current between instantaneous eigenstates $\ket{y(t)}$ and $\ket{x(t)}$, and $f_k^{xy}(t)$, like in the classical case, can be interpreted as a generalised thermodynamic force. In addition,  $\phi_k^{xy}(t)$ is the entropy flow associated with the jump $\hat{L}_k$ at time $t$ resulting in a probability current $j^{xy}_k(t)$ from state $\ket{y(t)}$ to $\ket{x(t)}$.
The rate equation for $p_x(t) $ may then be rewritten in terms of the probability currents between instantaneous eigenstates
\begin{equation}
\begin{split}
    \dt{p_x(t)} &= \sum_{k, y (y\neq x)}j_k^{xy}(t).
\end{split} 
\end{equation}
In section~\ref{sec: t_param} we have illustrated that the QFI with respect to the time parameter can always be split into an incoherent and a coherent contribution. Crucially, the incoherent contribution dependents solely on the instantaneous spectrum of the state. We may thus play the same game as in the classical case and relate the incoherent part of the QFI to thermodynamic observables.
{Again, we may express} $\dot{S}(t)$ in terms of the probability currents and further split it into the entropy production rate $\sigma$ and the entropic flow rate to the environment $\Phi$
\begin{equation}
    \begin{split}
    \dot{S}(t)  =& -\sum_{x,y, k , x\neq y} j_k^{xy}(t) \log(p_x(t))\\
    =& \frac{1}{2}\sum_{x,y, k } j_k^{xy}(t) \log\left(\frac{w^{xy}_k(t) p_y(t)}{w^{yx}_{k^\prime}(t) p_x(t)}\right) \\&- \frac{1}{2}\sum_{x,y, k } j_k^{xy}(t) \log\left(\frac{w^{xy}_k(t)}{w^{yx}_{k^\prime}(t)}\right).\\
    =&  \sigma(t) -\Phi(t),
\end{split}
\end{equation}
as shown in~\cite{van_vu_thermodynamic_2023}. In analogy to the classical result, the entropy production rate can be expressed as the current weighted average of the thermodynamic force $\sigma(t)= \langle \langle f(t) \rangle \rangle$, and the entropic flow is $\Phi(t) = \langle \langle \phi(t) \rangle \rangle$. \\
{We find that the incoherent contribution to the QFI $F_Q^{IC}(t)$ is given by}
\begin{equation}
    F_Q^{IC}(t) = \mathcal{B} - (\dot{\sigma}(t) -\dot{\Phi}(t)),
\end{equation}
where $\mathcal{B} = - \sum_x \ddot{p}_x \log p_x${, mirroring the classical result in~\cite{nicholson_nonequilibrium_2018}.
Therefore, it may be alternatively expressed as~\cite{ito_stochastic_2018}}
\begin{equation}
    F_Q^{IC}(t) = - \left \langle \left \langle \frac{\mathrm{d}f(t)}{\mathrm{d}t}\right \rangle\right \rangle + \left \langle \left \langle \frac{\mathrm{d}\phi}{\mathrm{d}t}\right\rangle\right \rangle.
\end{equation}
The above constitutes the first central result of our work.
We further observe that, as in the classical case~\cite{ito_stochastic_2018} discussed in section~\ref{sec: info_geo_thermo_classical}, the thermodynamic inequality $\left\langle\left\langle \frac{\mathrm{d}\phi(t)}{\mathrm{d}t} \right\rangle\right\rangle \geq \left\langle\left\langle \frac{\mathrm{d}f(t)}{\mathrm{d}t} \right\rangle\right\rangle$, which establishes the relationship between the rate of change of the thermodynamic force and the rate of environmental entropy change, also applies to the quantum regime.

Interestingly, despite it being finite and contributing to the statistical length, the coherent contribution to the QFI plays no role in this balance. Rather, like in the classical case, only the dynamics of the populations in the instantaneous eigenbasis are relevant.\\
In the following section, however, we derive an extension to a classical bound on the change in the rate of the von Neumann entropy in which the coherent contribution to the QFI indeed plays an explicit role.
\section{Geometric bound on the von Neumann entropy rate} 
\label{sec: bounding_Sdot}
In this section, we will first demonstrate that the classical uncertainty relation between the geometric uncertainty in the path in state space and the {time-averaged rate} of information change, established in~\cite{nicholson_nonequilibrium_2018} for classical systems, remains valid for quantum dynamics. {In fact, we find that the bound, for both classical and quantum systems, can be tightened by a factor of two.} We then make use of the decomposition of any QFI into a common metric-independent incoherent contribution and a metric-dependent coherent contribution, revised in section~\ref{sec: t_param}. The incoherent part is inherently linked to the system's entropy dynamics as shown in section~\ref{sec: info_geo_thermo_quantum}. This connection enables us to extend a classical bound on the change in the entropy rate, as described in~\cite{nicholson_nonequilibrium_2018}, to the quantum domain.

The time-averaged variance of the QFI, known as the geometric uncertainty, established in the classical case~\cite{nicholson_nonequilibrium_2018}, is given by
\begin{equation}
    \delta = 4 \frac{\mathcal{J}-\mathcal{L}^2}{T^2} = \mathrm{E}[F_Q] -  \mathrm{E}[\sqrt{F_Q}]^2\geq 0,
\end{equation}
{at any time $T$}, so that $0\leq T\leq \tau$. Here,  $\mathcal{J}$ and $\mathcal{L}$ denote the statistical divergence and statistical length accumulated up to time $T$, respectively.
The equation above captures the cumulative path deviation from the geodesic, which represents the trajectory of constant speed between the initial and final states. Since $\mathcal{J}-\mathcal{L}^2$ vanishes along the geodesic, where the QFI remains constant by definition, any non-zero difference indicates a deviation from this optimal path.
Because of the positivity of $\mathcal{J}$, $\mathcal{L}^2$ and $\delta$, one finds that
\begin{equation}
    \delta \leq \frac{4\mathcal{J}}{T^2}.
\end{equation}
Here it is helpful to rewrite the above inequality in terms of the time averaged Fisher information, which represents the time-averaged rate of information change, $\mathcal{I} = \mathrm{E}[F_Q] = \frac{1}{T} \int_{0}^{T}\mathrm{d}t F_Q$. The inequality then becomes
\begin{equation}
   \frac{\mathcal{I}}{\delta} \geq 1.
\end{equation}
The above takes the form of an uncertainty relation, indicating that lower uncertainty in the path comes at the cost of a lower time-averaged rate of information change, as argued in~\cite{nicholson_nonequilibrium_2018}. {Note, however, that the lower bound derived in~\cite{nicholson_nonequilibrium_2018} is looser than the bound presented here by a factor of $\frac{1}{2}$. This difference arises from the use of the inequality $\mathcal{J}-\mathcal{L}^2\leq 2\mathcal{J}$ in their argument, whereas our derivation relies on both $\mathcal{J}-\mathcal{L}^2\geq 0 $ and $\mathcal{L}^2\geq 0 $, leading to the tighter bound $\mathcal{J}-\mathcal{L}^2\leq \mathcal{J}$.} We find that this uncertainty relation holds for both classical and quantum dynamics.

{Saturation of the bound can be achieved in the long-time limit during a relaxation process to a steady state, where the QFI with respect to the time parameter vanishes by definition, as the state becomes constant:
        \begin{equation}
        \begin{split}
            \lim_{\tau\to \infty} \frac{\mathcal{I}}{\delta} &= \lim_{\tau\to \infty} \frac{1/\tau \int_0^\tau dt F_Q(t)}{1/\tau \int_0^\tau dt F_Q(t) - 1/\tau^2\left(\int_0^\tau dt \sqrt{F_Q(t)}\right)^2} \\
            &= \lim_{\tau\to \infty} \frac{ \int_0^\tau dt F_Q(t)}{\int_0^\tau dt F_Q(t) - \underbrace{ 1/\tau\left(\int_0^\tau dt \sqrt{F_Q(t)}\right)^2}_{\to 0}} = 1.
        \end{split}
        \end{equation}
This saturation behavior is discussed and demonstrated in the example of the quantum Mpemba effect in Sec.~\ref{sec: mpemba}.}
Building on the above uncertainty relation, we decompose the QFI into its incoherent and coherent components, $F_Q = F_Q^{IC} + F_Q^{C}$  and $F_Q^{IC}= \mathcal{B} - \ddot{S} $. By performing the integral $\int_{0}^{T}F_Q\mathrm{d}t  = \int_{0}^{T} \left(\mathcal{B}(t) + F_Q^C(t)  - \ddot{S}(t)\right)\mathrm{d}t$, we find that the difference in the entropy rate between the final and initial states is bounded from above
{
\begin{equation}
\label{eq: change_entropy_rate_bound}
\begin{split}
     \Delta \dot{S} &\leq \mathcal{C} - T\delta + \int_{0}^{T}\mathrm{d}t F_Q^C 
     ,\\
\end{split}
\end{equation}}
where $\Delta \dot{S} = \dot{S}(T) - \dot{S}(0)$ and $\mathcal{C} = \int_{0}^{T}\mathcal{B}(t)\mathrm{d}t $, which is the integrated difference between the the average local information rate and the bulk information rate, for $0\leq T \leq \tau$. 

The above bound constitutes the second key result of our work. Similar to the classical case~\cite{nicholson_nonequilibrium_2018}, the change in the entropy rate is constrained by the sum of the integrated difference between the average local information rate and the bulk information rate, as well as the geometric uncertainty. However, an additional non-negative quantum contribution emerges, which we identify as the geometric action associated with the coherent dynamics. {As briefly described in Sec.~\ref{sec: t_param}, the action is associated to the `energy' of the path due to its similarity with the kinetic energy integral in classical mechanics.} {It is worth pointing out that the first two terms vanish only in stationary states~\cite{nicholson_nonequilibrium_2018}, whereas the last term vanishes in stationary states or whenever the dynamics is purely incoherent. Therefore, their sum vanishes only in stationary states. }
Note that if one considers a driving protocol where the initial state is the fixed point of the initial instantaneous generator of the dynamics, such that $\dot{S}(0)=0$, the above serves as a bound on the instantaneous rate of entropy change $\dot{S}(T)$ at any later time $T${, which indicates that the state is out of equilibrium}.
We can further exploit the relationship between the incoherent contribution to the QFI and the entropic acceleration in Markovian open system dynamics obeying local detailed balance. In cases where the heat flow between the system and its environment vanishes, the resulting bound applies to the entropy production. Conversely, if the entropy production is zero, the bound constrains the entropic flow. 

It is also worth noting that our bound applies irrespective of the chosen QFI metric. {However, it can be shown that since  $- \delta T + \int_{0}^{T}\mathrm{d}t F_Q^C = 4\mathcal{L}_f^2/T - \int_{0}^{T}\mathrm{d}t F_Q^{IC}$, it is minimized for the SLD QFI, resulting in the tightest bound. }

\section{Quantum Mpemba effect}
\label{sec: mpemba}
{We now apply the information-geometric framework developed in the previous sections to analyze the (thermo-)dynamics underlying a quantum Mpemba effect. This effect is manifest during the thermal relaxation of a qubit interacting with a thermal bosonic bath and was recently explored in~\cite{moroder_thermodynamics_2024}.}
Historically, the Mpemba effect~\cite{mpemba_cool_1969, kell_freezing_1969} describes the following phenomenon: consider two identical buckets of liquid. Both are at thermal equilibrium, but with respect to environments at different temperatures, $T_h$ and $T_c$, with $T_h>T_c$. Then, these buckets are placed in a third environment with a lower temperature $T_h>T_c>T^*$, and are allowed to thermalise. A Mpemba effect is said to be observed if there exists a time $t_M$ after which the temperature of the initially hotter liquid is lower than that of the initially colder liquid at all later times. The effect has stimulated a lot of theoretical and experimental interest \cite{hu_conformation_2018, ahn_experimental_2016, chaddah_overtaking_2010, greaney_mpemba-like_2011, lasanta_when_2017, keller_quenches_2018, turkeshi_arxiv_2024}, however, a (universal) physical origin remains elusive and consensus on whether the effect exists in the first place has not yet been reached~\cite{burridge_questioning_2016, bechhoefer_fresh_2021}. More recently, Mpemba-like effects have been predicted and experimentally observed also in the quantum domain~\cite{ares_entanglement_2023, murciano_entanglement_2024, yamashika_entanglement_2024, liu_symmetry_2024, joshi_observing_2024}, primarily focused on symmetry-restoration dynamics in closed quantum systems. 

We address a particular version occurring in the {thermalisation} behaviour of Markovian open quantum systems, recently put forward by Moroder et al.~\cite{moroder_thermodynamics_2024}. The authors used the concept of non-equilibrium free energy $F_\mathrm{neq}$ to understand the thermodynamics of the effect. This is defined as
\begin{equation}
    F_{\mathrm{neq}} (\hat{\sigma}) = \beta^{-1} D(\hat{\sigma} \| \hat{\tau}_\beta) + F_{\mathrm{eq}},
\end{equation}
with equilibrium free energy $F_{\mathrm{eq}} = -\beta^{-1} \log Z_\beta $, partition function $Z_\beta = \mathrm{Tr}\left[\exp(-\beta \hat{H})\right]$ and $D(\hat{\sigma} \| \hat{\tau}_\beta)  = \mathrm{Tr}\left[\hat{\sigma} \left(\log \hat{\sigma}  - \log \hat{\tau}_\beta \right)\right]$
is the quantum relative entropy between the relaxing state and the thermal fixed point {$\hat{\tau}_\beta = \exp(-\beta \hat{H})/Z_\beta$, with respect to the system Hamiltonian $\hat{H}$}. The authors in \cite{moroder_thermodynamics_2024} defined a Mpemba effect to occur when a $\hat{\rho}_0'$ with higher initial non-equilibrium free energy $F_\mathrm{neq}$ compared to the reference state $\hat{\rho}_0$, relaxes to thermal equilibrium faster. In particular, there exists a time $t_M$ after which {$F_\mathrm{neq}(\hat{\rho}(t)) > F_\mathrm{neq}(\hat{\rho}'(t))$} for all $t > t_M$, indicating a crossing of non-equilibrium free energies along their respective relaxation paths. \\
\begin{figure}[t]
\begin{center}
\includegraphics[width=\linewidth]{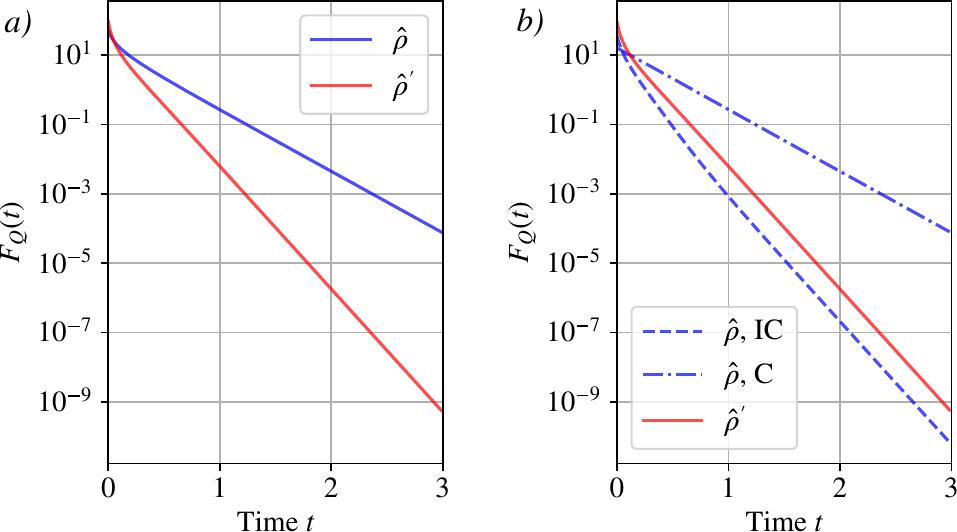}
  \caption{a) Time-evolution of the total SLD QFI of the reference state $\hat{\rho}$ (blue) and the rotated state $\hat{\rho}^\prime$ (red). b) The rotated state evolves classically (red), therefore the quantum Fisher information reduces to the incoherent contribution. The reference state, however, is coherent in the energy eigenbasis and thus has both incoherent (dashed blue) and coherent (dash-dotted) contributions. We find that the incoherent contribution decays exponentially faster than the coherent contribution.}
 \label{fig: fisher_info}
\end{center}
\end{figure}
\begin{figure}[t]
\begin{center}
\includegraphics[width=\linewidth]{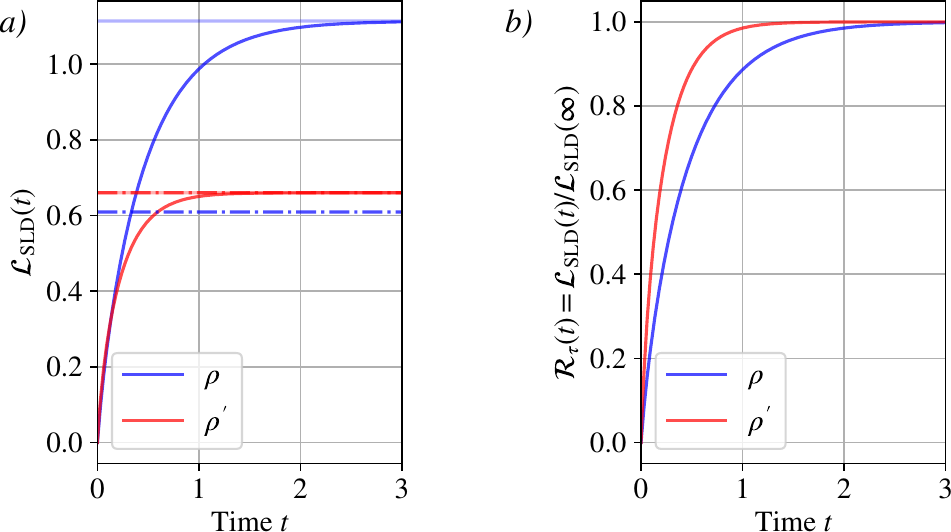}
  \caption{a) Statistical distance (SLD) traced by the path of the reference state $\hat{\rho}$ (blue solid) and the rotated state $\hat{\rho}^\prime$ (red solid) in state space. The semi-transparent lines indicate the distance traced by the respective paths in the infinite time limit, indicating that sufficient thermalisation is reached within the chosen time period. The statistical distance for the rotated state is shorter than for the reference state. However, the SLD geodesic distance between the reference state and the thermal state (blue dash-dotted) is shorter than that between the rotated state and the thermal state (red dash-dotted), which coincides with the statistical distance.   b) Furthermore, we find that the rotated state (red) completes the path at a faster rate as shown by the higher ratio of completion at all times.}
 \label{fig: ratio of completion}
\end{center}
\end{figure}
\begin{figure}[t]
\begin{center}
\includegraphics[width=\linewidth]{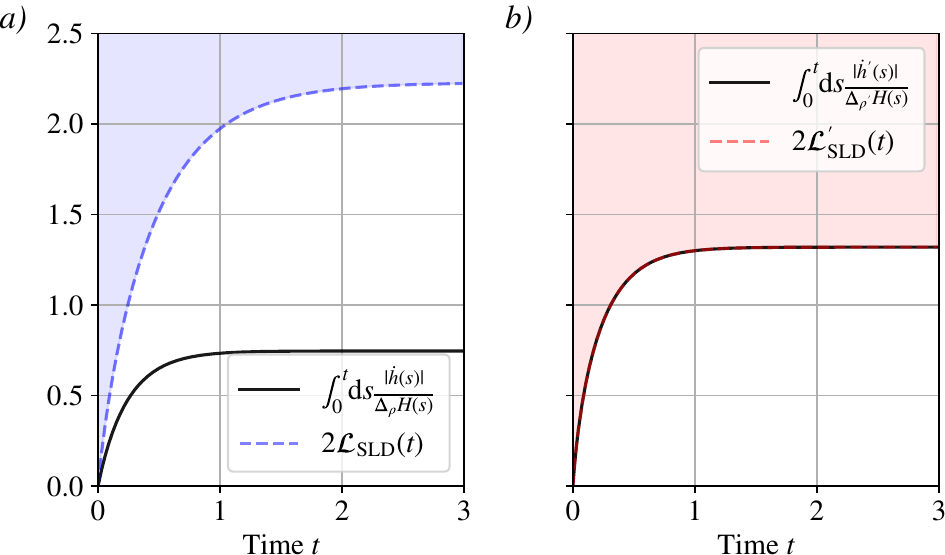}
  \caption{The path-dependent time-integrated ratio between the heat current and the standard deviation of the Hamiltonian with respect to the time-evolving state is upper bounded by twice the statistical distance associated with the path of the time-evolving density matrix in state space, shown for a) the reference state $\hat{\rho}$ and b) the rotated state $\hat{\rho}^\prime$. The path traced by the rotated state coincides with that of the geodesic (even though it is not traced at constant speed). Therefore, the bound set by the statistical distance coincides with the generally tightest one provided by the length of the geodesic. The bound is saturated because the state evolves purely incoherently.}
 \label{fig: integrated_bounds}
\end{center}
\end{figure}
Consider a qubit weakly coupled to a thermal bosonic environment at inverse temperature $\beta$. Interactions between the ground state $\ket{g}$ and the excited state $\ket{e}$ separated by energy $\epsilon$ ($\hbar=k_\mathrm{B}=1$), with Hamiltonian $\hat{H} = \frac{\epsilon}{2} \hat{\sigma}_z$, are facilitated by interactions with a bosonic environment. The transitions occur with rates $\gamma_+ = \gamma f_B(\epsilon, \beta)$ and $\gamma_- = \gamma (1+f_B(\epsilon, \beta))$, with Bose-Einstein occupation $f_B(\epsilon, \beta) = \left[\exp(\beta \epsilon)-1\right]^{-1}$ {and coupling strength $\gamma$}. 
The GKSL master equation that describes the resulting reduced dynamics of the qubit is given by
\begin{equation}
\label{eq: master_eq}
    \dt{\hat{\rho}(t)} = -i\left[\hat{H}, \hat{\rho}(t)\right] + \mathcal{D}\left[\hat{L}_+\right]\hat{\rho}(t) +  \mathcal{D}\left[\hat{L}_-\right]\hat{\rho}(t),
\end{equation}
with jump operators accounting for absorption and emission events, $\hat{L}_+ = \sqrt{\gamma_+} \ket{e}\bra{g}$ and $\hat{L}_- = \sqrt{\gamma_-} \ket{g}\bra{e}$,
and the super-operator $\mathcal{D}\left[\hat{L}\right]\hat{\rho} = \hat{L}\hat{\rho} \hat{L}^\dagger - \frac{1}{2}\left\lbrace \hat{L}^\dagger \hat{L}, \hat{\rho} \right\rbrace$. \\
The qubit is initialised to an { arbitrary} mixed state, serving as the reference state $\hat{\rho}_0$. It is parameterised as $
    \hat{\rho}_0 = \frac{1}{2}\left(1 + \vec r \cdot \vec \sigma \right)$,
where $\vec \sigma = (\hat{\sigma}_x, \hat{\sigma}_y, \hat{\sigma}_z)^\mathrm{T}$ and $\vec r = (r_x, r_y, r_z)^\mathrm{T}$. We restrict to reference initial states with coherences in the energy eigenbasis so that at least $r_x$ or $r_y$ are non-zero. {A} unitarily rotated state, diagonal in the energy eigenbasis, is $\hat{\rho}^\prime_0 = \mathrm{diag}(\lambda_1, \lambda_2)$,
where $\lambda_1$ and $\lambda_2$ are the eigenvalues of $\hat{\rho}$, so that $\lambda_1 \geq \lambda_2$. Importantly, {this} choice of $\hat{\rho}_0^\prime$ ensures that $ F_{\mathrm{neq}} (\hat{\rho}_0^\prime)\geq F_{\mathrm{neq}} (\hat{\rho}_0)$.

The numerical results, which we will describe shortly, are obtained for the reference state and the rotated state, specified by the vectors $\vec r = (-0.41760,-0.60647, 0.47879)$ and  $\vec{r}' = (0,0, 0.87836)$, respectively. { Additional parameters (with $\hbar=k_\mathrm{B}=1$) that determine the set-up are the coupling strength to the bosonic reservoir $\gamma = 1$, the temperature $T = 10$ and the Hamiltonian level splitting {$\epsilon = 5$.}}

Note that, in this context, the interaction with the environment cannot generate coherence in the energy eigenbasis. Thus, under dynamics induced by the GKSL master equation (Eq.~\eqref{eq: master_eq}), the rotated state will remain diagonal in the eigenbasis of the Hamiltonian at all later times. We now remember that a general time evolution can be understood as the combination of time-evolving eigenvalues, represented by $\chi(t)$ in a time evolving eigenbasis, expressed in $U_t$ (see section~\ref{sec: t_param}). But since the eigenbasis remains static, $U_t$ is the identity, and {the dynamics are fully captured} by the time-evolving eigenvalues. Therefore, the QFI with respect to time reduces to its incoherent contribution - the classical Fisher information for the probability distribution constructed from the time-evolving eigenvalues. The reference state, however, initially has coherence in the energy eigenbasis by construction, and therefore the QFI has both an incoherent and a coherent contribution, as shown in Fig.~\ref{fig: fisher_info}b). 

As discussed in section~\ref{sec: QFI}, $\sqrt{F_Q(t)}$ quantifies the speed at which a state moves along its path in state space. At equilibrium, the state is stationary, and therefore, the QFI vanishes. Our findings reveal that at short times, the rotated state (with initially higher non-equilibrium free energy) evolves more rapidly towards equilibrium. In contrast, the coherent reference state reaches the equilibrium state much later, as indicated by the slower decrease in the total QFI, as shown in Fig.~\ref{fig: fisher_info}a).
This difference can be understood by examining the incoherent and coherent contributions separately. The coherent contribution to the QFI of the reference state, {due to Hamiltonian dynamics}, decays exponentially slower than the incoherent contribution {stemming from  the dissipative dynamics}, and thus dictates the relaxation timescale. Since the dynamics of populations and coherences are decoupled~\cite{davies_generators_1979, roga_davies_2010}, the decay rates of the incoherent contributions to the QFI are identical for both the reference and rotated states (Fig.~\ref{fig: fisher_info}).\\
Fig.~\ref{fig: ratio of completion} shows that the SLD statistical distance of the path traced by the time-evolving incoherent initial state (light red), as it evolves to the thermal state, is significantly smaller compared to that of the coherent initial state (light blue)., i.e. $\mathcal{L}^\infty_\mathrm{SLD}(\hat{\rho}_0') < \mathcal{L}^\infty_\mathrm{SLD}(\hat{\rho}_0)$, where 
\begin{equation}
    \mathcal{L}_\infty(\hat{\rho}_0) = \lim_{t\to \infty}\int_0^t  \sqrt{F_Q(\hat{\rho}(t'))} \mathrm{d}t'.
\end{equation}
Interestingly, for the SLD geodesic lengths (see Eq.~\eqref{eq: geodesic_length}) we find the opposite to be true (dashed lines), as shown in Fig.~\ref{fig: ratio of completion}a).
Although the rotated state does not evolve along a geodesic - meaning the QFI is not constant along the path - the length of its path matches that of a geodesic at each instant, which describes a strictly monotonic trajectory along the $z$-axis of the Bloch sphere. Additionally, we find that the ratio of completion, indicating the fraction of the statistical distance traced as a function of time is higher for the rotated state than for the reference state $\mathcal{R}^\prime_\tau(t) = \mathcal{L}_\mathrm{SLD}(\hat{\rho}'(t))/\mathcal{L}_\mathrm{SLD}(\hat{\rho}'(\tau)> \mathcal{R}_\tau(t) = \mathcal{L}_\mathrm{SLD}(\hat{\rho}(t))/\mathcal{L}_\mathrm{SLD}(\hat{\rho}(\tau))$, as shown in Fig.~\ref{fig: ratio of completion} b), indicating further that the rotated state reaches equilibrium faster than the reference state. \\ 
\begin{figure}[t]
\begin{center}
\includegraphics[width=\linewidth]{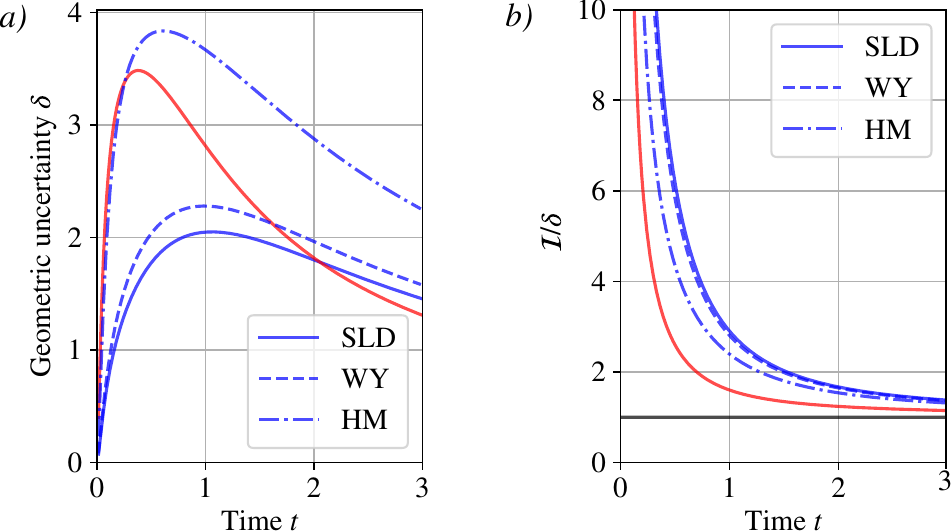}
  \caption{a) Geometric uncertainty $\delta$ about the path of the rotated state $\hat{\rho}^\prime$ (red). We compute the geometric uncertainty of the time-evolving reference state $\hat{\rho}$ using the SLD, WY and HM metric (blue). The SLD yields the smallest geometric uncertainty. Deviations from the geodesic connecting the reference state and rotated state to the thermal state, respectively, are accumulated at short times.  b) Trade-off between the time-averaged QFI and the geometric uncertainty. }
 \label{fig: geometric uncertainty}
\end{center}
\end{figure}
We now shift focus to a geometric bound on the time-evolution of observables and their fluctuations.
The bound we derived in section~\ref{sec: bounding_Sdot} illustrates the possible benefit of choosing one QFI over another. Another advantage of the SLD is its direct link to the time-derivative of the density matrix. 
An interesting geometric consequence of this connection {for arbitrary observables $\hat{O}$ was first established in~\cite{garcia-pintos_unifying_2022} (recently generalised also to other quantum Fisher metrics in~\cite{bringewatt_arxiv_2024})},
\begin{equation}
    \int_0^\tau \frac{\left\lvert\dot{o}(t)\right\rvert}{\Delta_{\hat{\rho}_t} \hat{O}}\mathrm{d}t\leq \int_0^\tau \sqrt{F_Q} \mathrm{d}t = 2\mathcal{L}_\mathrm{SLD},
\end{equation}
expressing that the time-integrated ratio between the change in the observable expectation value due to state changes, $\left\lvert \dot{o}(t)\right\rvert = \left\lvert\mathrm{Tr}\left[\hat{O}\frac{\mathrm{d}\hat{\rho}_t}{\mathrm{d}t}\right]\right\rvert$, and the standard deviation of the observable with respect to the instantaneous state $\hat{\rho}_t$, $\Delta_{\hat{\rho}_t} \hat{O} = \sqrt{\mathrm{Tr}\left[\hat{\rho}_t \hat{O}^2\right] - \mathrm{Tr}\left[\hat{\rho}_t \hat{O}\right]^2}$, is upper bounded by twice the SLD statistical distance $\mathcal{L}_\mathrm{SLD}$. Note that since $\mathcal{L}_\mathrm{SLD}\geq\mathcal{L}_\mathrm{SLD}^\mathrm{geo}$ (see Eq.~\eqref{eq: geodesic_length}), the  ratio integrated over a geodesic path yields the lowest bound. 

Here, we choose the observable of interest to be the Hamiltonian $\hat{H}$, as its changes in the expectation value due to state changes is precisely the heat current, which, for consistency with previously introduced notation, we denote as $\left\lvert\dot{h}(t)\right\rvert$. In Fig.~\ref{fig: integrated_bounds}, we show the corresponding integrated ratio (black line) in comparison to its bound defined by the statistical distance (blue and red dashed lines). The colored regions represent the areas excluded by these bounds. We first note that since the statistical distance between the initial rotated state $\rho^\prime$ and the time-evolved state  $\rho^\prime (t)$ coincides with the geodesic length between the two at all times, they provide the identical bound (Fig.~\ref{fig: integrated_bounds}b)). {We find further that this bound is saturated at all times, a consequence of the fact that along the incoherent trajectory, the covariance between the logarithmic derivative $\hat{L}_\text{SLD}$ and the Hamiltonian $\hat{H}$ factorises into the the product of their respective standard deviation with respect to $\hat{\rho}'_t$, so that $|\dot{h}(t)| = \text{cov}_{\hat{\rho}'_t}(\hat{H}, \hat{L}_\text{SLD}) = \Delta_{\hat{\rho}'_t} \hat{H} \Delta_{\hat{\rho}'_t}\hat{L}_\text{SLD} = \Delta_{\hat{\rho}'_t}\hat{H} \sqrt{F_Q(t)}$, as discussed in detail in~\cite{garcia-pintos_unifying_2022}.} Clearly, this is not the case for the initially coherent reference state (Fig.~\ref{fig: integrated_bounds}a)). 

We now focus on the the geometric uncertainty, which for the evolution of the coherent reference state, is metric-dependent. Generally, we find that the cumulative deviation from the respective geodesic paths of the two initial states, is accumulated at short times during the relaxation, as shown in Fig.~\ref{fig: geometric uncertainty}a). Qualitatively, we find further that the geometric uncertainty is initially higher for the incoherent state  irrespective of the chosen QFI metric. Interestingly, the time at which the geometric uncertainty in the rotated state crosses those of the reference state, however, depends on the chosen metric.

Finally, we examine the trade-off relation between the the time-averaged Fisher information $\mathcal{I}$ and the geometric uncertainty $\delta$. It expresses that lower uncertainty in the path comes at the cost of lower time-averaged rate of information change, as discussed in section~\ref{sec: bounding_Sdot}.  Initially, all paths deviate strongly from the geodesic. The bound tells us that the high geometric uncertainty (and thus a high time-variance of the speed of evolution) has to be compensated by a sufficiently large time-averaged speed. We find that this ratio is consistently lower, and thus closer to the bound, during the relaxation of the rotated state at all times, as shown in Fig.~\ref{fig: geometric uncertainty}b). {At long times, the bound approaches saturation because the state nears the thermal steady state of the dynamics, where $F_Q$ vanishes due to the state's time invariance, as discussed in Sec.~\ref{sec: bounding_Sdot}.}
\section{Conclusion}
 \label{sec: conclusion}
{In our work, we present a fully quantum mechanical derivation that connects information-geometric quantities, defined via the quantum Fisher information (QFI) with respect to the time parameter, to quantum thermodynamics within the framework of stochastic processes. This enables us to extend classical results on entropy dynamics in nonequilibrium systems to the quantum domain.}
Specifically, we demonstrate that for open quantum systems governed by GKSL dynamics, any QFI is connected to the accelerations of entropy production and entropic flow. To this end, we leverage the fact that any QFI about the time parameter can be decomposed into coherent and incoherent contributions, and we find that the entropic acceleration is expressible solely in terms of the incoherent part. We {tighten} the classical uncertainty relation between the geometric uncertainty in the path in state space and the time-integrated rate of information change, and show that it is valid also for quantum dynamics. Further, we demonstrate that the classical bound on changes in the von Neumann entropy rate can be extended to the quantum domain through the addition of the non-negative geometric action associated with the coherent dynamics. {While this bound a priori holds for any QFI, we find that it is tightest for the SLD QFI.} 

Finally, we show that the thermodynamic quantum Mpemba effect, recently reported in~\cite{moroder_thermodynamics_2024}, can be understood through the analysis of geometric quantities, such as the ratio of completion and deviations from the geodesic path. Given the inherently general nature of the information-geometric framework, and its demonstrated applicability to quantum dynamics --- while maintaining connections to thermodynamic quantities as shown in our work and others~\cite{garcia-pintos_unifying_2022, tejero_asymmetries_2024, bringewatt_arxiv_2024} --- we aim to use it in future research to uncover common underlying mechanisms driving various anomalous dynamics, such as Mpemba-like effects, in both closed and open quantum systems.

 \begin{acknowledgements}
 We thank Ronnie Kosloff, Ivan Medina and Artur Lacerda for fruitful discussions and feedback. LPB and JG acknowledge Research Ireland for support through the Frontiers for the Future project. JG is funded by a Science Foundation Ireland-Royal Society University Research Fellowship. JG thanks the Yukawa Institute for Theoretical Physics at Kyoto University Japan where some of this work was completed during Frontiers in Non-equilibrium Physics 2024.
\end{acknowledgements}

\bibliography{
misc_refs.bib, references_info_geo.bib}

\bibliographystyle{apsrev4-2}
\clearpage
\appendix

\end{document}